\documentclass[aps,twocolumn,groupedaddress,showpacs]{revtex4}
\usepackage{graphicx}
\usepackage{color}
\usepackage{amsmath}
\usepackage{bm}
\usepackage{txfonts}
\usepackage{ulem}

\begin{document}

\title{
Molecular kinetic analysis of a local equilibrium Carnot cycle
}

\author{Yuki Izumida}
\thanks{Present address: Department of Complex Systems Science, Graduate School of Informatics, Nagoya University, Nagoya 464-8601, Japan}
\email{izumida@i.nagoya-u.ac.jp}
\affiliation{Department of Complex Systems Science, Graduate School of Information Science, Nagoya University, Nagoya 464-8601, Japan}
\author{Koji Okuda}
\affiliation{Division of Physics, Hokkaido University, Sapporo 060-0810, Japan}

\begin{abstract}
We identify a velocity distribution function of ideal gas particles that is compatible with the local equilibrium assumption and the fundamental thermodynamic relation satisfying the endoreversibility.
We find that this distribution is a Maxwell--Boltzmann distribution with a spatially uniform temperature and a spatially varying local center-of-mass velocity.
We construct the local equilibrium Carnot cycle of an ideal gas, based on this distribution, and show that the efficiency of the present cycle is given by the
endoreversible Carnot efficiency using the molecular kinetic temperatures of the gas. 
We also obtain an analytic expression of the efficiency at maximum power of our cycle under a small temperature difference.
Our theory is also confirmed by a molecular dynamics simulation.
\end{abstract}
\pacs{05.70.Ln}

\maketitle

\section{Introduction}\label{Introduction}
Global equilibrium between the working substance and the heat reservoir as the reversibility condition is essential for the thermodynamic cycle of heat engines to attain the maximum efficiency (Carnot efficiency)~\cite{C,Reif}.
Denoting by $Q_h$ ($Q_c$) the heat from the hot (cold) heat reservoir with the temperature $T_h^{\rm R}$ ($T_c^{\rm R}$) ($T_h^{\rm R}>T_c^{\rm R}$)
during the isothermal processes, 
we can express the global equilibrium as the Clausius equality applied to the Carnot cycle:
\begin{eqnarray}
\frac{Q_h}{T_h^{\rm R}}+\frac{Q_c}{T_c^{\rm R}}=0,
\end{eqnarray}
from which the efficiency $\eta\equiv \frac{W}{Q_h}$ of the heat-energy conversion into work $W\equiv Q_h+Q_c$ 
is given by the Carnot value $1-\frac{T_c^{\rm R}}{T_h^{\rm R}}\equiv \eta_{\rm C}$.
For this global equilibrium to hold, the heat engine should run along the cycle infinitely slowly (quasistatic limit) and hence output zero power (work per unit time).

Curzon and Ahlborn (CA)~\cite{CA}
considered the efficiency at maximum power $\eta^*$ as a more practical figure of merit (The same subject was also considered by some authors even earlier.
See~\cite{VLF,MP} for historical perspectives on the origin of the efficiency at maximum power and references therein). 
CA assumed that their heat engine cycle (CA cycle) satisfies the Fourier's law of heat transport and the so-called endoreversibility condition~\cite{R}, which is written explicitly for a cycle as
\begin{eqnarray}
\frac{Q_h}{T_h}+\frac{Q_c}{T_c}=0,\label{eq.endoreversibility1}
\end{eqnarray}
where $T_h$ ($T_c$) is the well-defined temperature of the working substance in contact with the hot (cold) heat reservoir during the isothermal process at a finite rate.
From this, the efficiency of the CA cycle is given by the temperatures of the working substance as
\begin{eqnarray}
\eta=1-\frac{T_c}{T_h},\label{eq.endo_effi}
\end{eqnarray}
which we call the endoreversible Carnot efficiency.
This suggests that the efficiency of the endoreversible heat engine is still expressed by the 
Carnot-like expression, depending only on the temperatures of the working substance.
CA showed that Eq.~(\ref{eq.endo_effi}) at the maximum power becomes
\begin{eqnarray}
\eta^*=1-\sqrt{\frac{T_c^{\rm R}}{T_h^{\rm R}}}=1-\sqrt{1-\eta_{\rm C}}=\frac{\eta_{\rm C}}{2}+\frac{\eta_{\rm C}^2}{8}+O(\eta_{\rm C}^3),\label{eq.ca}
\end{eqnarray}
which we call the CA efficiency. 
This result gave birth to the field of finite-time thermodynamics~\cite{B,SNSAL,BKSST} that studies various thermodynamic systems performing finite-time transformations based on the endoreversibility.
Since the universality of Eq.~(\ref{eq.ca}) was addressed in~\cite{VB} based on linear irreversible thermodynamics, 
the efficiency at maximum power has been investigated as a fundamental problem in nonequilibrium thermodynamics~\cite{CH,SS2008,IO2008,IO2009,ELV,EKLV,IO2012,ST,WZ,HNE,PDCV,SH,BSS,PV,CPV}.

In a recent paper~\cite{IO2015}, the present authors 
showed that a physical origin of the endoreversibility Eq.~(\ref{eq.endoreversibility1}), which is usually simply assumed in finite-time thermodynamics, can be attributed to a special case of a local equilibrium assumption
(see also~\cite{R,WZ} for similar ideas).
Here, we refer to the local equilibrium assumption as an assumption where a total system 
is not in a global equilibrium state sharing the same intensive thermodynamic variables, 
while each partial system is in an equilibrium state with locally-defined thermodynamic variables~\cite{GM}.
The endoreversibility condition can then be regarded as
the special case of this local equilibrium assumption applied to the heat engine
constituted with the working substance and the heat reservoir:
The whole working substance itself is assumed to be in a local equilibrium state with the well-defined temperature $T$ without spatial variation,
where this temperature is different from that of the heat reservoir in a local equilibrium state,
while the global equilibrium between them is violated.
In this case, the following fundamental thermodynamic relation holds for the thermodynamic variables of the working substance,
\begin{eqnarray}
dU=TdS-pdV,\label{eq.ftr}
\end{eqnarray}
where $S$, $U$, $p$, and $V$ are the entropy, internal energy, pressure, and volume of the working substance, respectively.
Indeed, it can be shown that the endoreversibility condition Eq.~(\ref{eq.endoreversibility1}) holds automatically by applying the following closed-cycle condition to the cycle with constant temperatures during the isothermal processes
as the CA cycle:
\begin{eqnarray}
\oint dS=\oint \frac{dQ}{T}=0.
\end{eqnarray}
In this sense, we may say that the local equilibrium is an essential feature of the CA cycle as the endoreversible heat engine 
in such a manner that the global equilibrium is an essential feature of the Carnot cycle as the reversible heat engine~\cite{IO2015}.
However, how such a macroscopic and phenomenological description using the fundamental thermodynamic relation in a finite-time process can be established 
from a statistical mechanics point of view using a state distribution of the working substance is still not obvious, which would be of crucial importance to strengthen the foundation of finite-time thermodynamics.

In the present paper, from a molecular kinetic analysis, we identify a velocity distribution of ideal gas particles 
as the simplest case of the working substance that is consistent with the local equilibrium assumption and the fundamental thermodynamic relation Eq.~(\ref{eq.ftr}) satisfying the endoreversibility. 
Based on this distribution, 
we construct a local equilibrium Carnot cycle and study the efficiency at maximum power of our cycle by comparing it to the CA efficiency.
We also perform a molecular dynamics simulation to confirm the validity of our theory.

The rest of the paper is organized as follows.
In Sec.~\ref{Molecular kinetic model}, as a preparation, we introduce our molecular kinetic model of an ideal gas system in a cylinder with a moving piston, and derive the velocity distribution
of the gas particles. 
In Sec.~\ref{Local equilibrium Carnot cycle}, we construct our local equilibrium Carnot cycle based on the preparation in Sec.~\ref{Molecular kinetic model},
and study the efficiency at maximum power.
The molecular dynamics simulation is also given in this section.
We discuss and summarize the present paper in Sec.~\ref{Discussion and summary}.

\section{Molecular kinetic model}\label{Molecular kinetic model}
\begin{figure}
\includegraphics[scale=0.75]{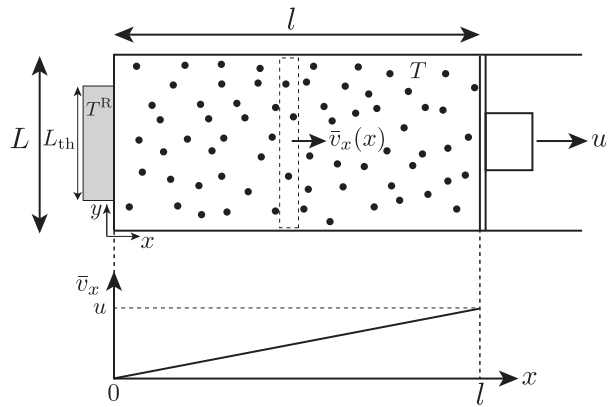}
\caption{Schematic illustration of $2$D ideal gas particles confined in a rectangular-shaped cylinder $l \times L$ with a piston on the head at $x=l$.
The piston moves at a constant velocity $u=\frac{dl}{dt}$.
The thermal wall with length $L_{\rm th}$ that mimics the interaction with the heat reservoir is set on the bottom of the cylinder at $x=0$.
The local center-of-mass $x$-velocity of the particles $\bar{v}_x(x)$ is shown to change linearly from $0$ at $x=0$ (bottom of the cylinder) to $u$ at $x=l$ (moving piston).
}\label{piston}
\end{figure}

\subsection{Ideal gas system in cylinder with moving piston}\label{Ideal gas system in cylinder with moving piston}
As a preparation for constructing our local equilibrium Carnot cycle, we first develop the molecular kinetics of the working substance in a cylinder with a moving piston.
We assume a two-dimensional (2D) ideal gas as the working substance for simplicity and assume that the temperature $T$ of the gas can be defined uniquely and the density of the gas is always uniform without spatial variation.
Imagine that $N$ ideal gas particles with mass $m$ are in a rectangular-shaped cylinder with dimensions $l\times L$ 
(Fig.~\ref{piston}).
At the bottom of the cylinder ($x=0$) is a thermal wall with length $L_{\rm th}$,
which realizes contact with the heat reservoir at the temperature $T^{\rm R}$ during an isothermal process. 
When a particle with velocity ${\bm v}=(v_x,v_y)$ collides with the thermal wall, its velocity stochastically changes to ${\bm v}'=(v_x',v_y')$ according to a normalized probability distribution~\cite{TTKB} (Maxwell boundary condition~\cite{K}),
\begin{eqnarray}
f_{\rm th}({\bm v}')=\frac{1}{\sqrt{2\pi}}\left(\frac{m}{k_{\rm B}T^{\rm R}}\right)^{3/2} v_x' \exp \left(-\frac{m({v_x'}^2+{v_y'}^2)}{2k_{\rm B}T^{\rm R}}\right),
\label{eq.distri_thermal}
\end{eqnarray}
where $0< v_x' < \infty$ and $-\infty < v_y' < \infty$.
This reflecting rule ensures that the temperature of the static gas becomes $T^{\rm R}$ (see also~\cite{CLL,BCM} and references therein for different types of thermal walls).
The heat flowing into the working substance per collision is calculated as the kinetic energy change before and after the collision, given by 
\begin{eqnarray}
\frac{m(|{\bm v}'|^2-|{\bm v}|^2)}{2}.\label{eq.micro_heat}
\end{eqnarray}
At the top of the cylinder ($x=l$) is a moving piston. When a particle with ${\bm v}$ collides with the piston moving with the constant velocity $u\equiv \frac{dl}{dt}$, where $t$ is the time, 
the particle velocity changes as ${\bm v}\to {\bm v}'=(-v_x+2u, v_y)$, where the mass of the
piston is assumed to be sufficiently larger than that of the particle and the collision is perfectly elastic.
The work done on the piston per collision is calculated in the same way as the heat in Eq.~(\ref{eq.micro_heat}), as follows:
\begin{eqnarray}
-\frac{m(|{\bm v}'|^2-|{\bm v}|^2)}{2}=2mu(v_x-u).\label{eq.micro_work}
\end{eqnarray}
Any particle collision with other parts of the cylinder and the particle--particle collisions are assumed to be perfectly elastic.

Here, we note that by our assumption, the gas must relax to a local equilibrium with a constant temperature $T(u)$, depending on $u$, much faster than the global equilibrium between the gas and the heat reservoir is realized. 
This is justified under the assumption of weak coupling between the working substance and the heat reservoir, where 
the time scale of the
equilibration inside the gas is much faster than the time scale of the energy exchange between the gas and the heat reservoir.
With this separation of the time scales, we can regard that the equilibration process of the gas into the local equilibrium state with the uniform temperature
by heat conduction inside the gas is instantaneous, 
and that the dynamics of the gas during the isothermal process is reduced to that of the temperature.
This would be realized in the case of a thermal wall with a sufficiently short length as $L_{\rm th}\ll L$ in the present setup~\cite{IO2008},
where the collision frequency with the thermal wall becomes much lower than that of the interparticle collisions.
In addition, 
our ideal gas should be precisely regarded as a ``weakly interacting nearly ideal gas," meaning that
the equilibration inside the gas is caused by interparticle collisions~\cite{IO2008}.

\subsection{Velocity distribution with local center-of-mass velocity}\label{sec.local_vel}
The local center-of-mass $x$ velocity $\bar{v}_x (x)$ of the particles located at position $\bm x=(x,y)$ 
can be uniquely determined according to the following argument:
Let us consider that the length of the cylinder changes as $l'=l+u\Delta t=l\left(1+\frac{u}{l}\Delta t\right)$ after an infinitesimal time duration $\Delta t$. 
We also consider a partial system $x\times L$ inside the cylinder $l\times L$, 
where the $x$-length of the partial system also changes as $x'=x+\bar{v}_x(x)\Delta t$ with the local center-of-mass $x$-velocity $\bar{v}_x(x)$ 
during $\Delta t$.
Before the displacement, the density of the entire system agrees with the density of the partial system with its particle number $N_x$ as $\frac{N}{lL}=\frac{N_x}{xL}$ from the uniformity of the density over the entire system.
Assuming that the particle number of the partial system after the displacement $N_x'$ is conserved as $N_x'=N_x$, and using the uniformity of the density over the entire system after the displacement as $\frac{N}{l'L}=\frac{N_x'}{x'L}$,
we can obtain the relation $\frac{x'}{l'}=\frac{x}{l}$.
We then obtain 
\begin{eqnarray}
x'=\frac{l'}{l}x=x+\frac{x}{l}u\Delta t,
\end{eqnarray}
which identifies $\bar{v}_x(x)$ as 
\begin{eqnarray}
\bar{v}_x(x)=\frac{x}{l}u \ \ (0\le x \le l).\label{eq.local_velocity}
\end{eqnarray}
We can also validate Eq.~(\ref{eq.local_velocity}) based on 
the inviscid Navier-Stokes equations 
(see the Appendix~\ref{appendix}).

If we look at the particle velocities at position ${\bm x}$ in the moving frame with the local center-of-mass velocity $\bar{\bm v}=(\bar{v}_x(x),0)$, that is, under a variable transformation ${\bm v} \to \tilde{\bm v}
\equiv {\bm v}-\bar{\bm v}=
(v_x-\bar{v}_x(x), v_y)$,
the velocity distribution
measured in this frame should be equal to the usual Maxwell--Boltzmann distribution with $T$ as
\begin{eqnarray}
f_{\rm MB}(\tilde{{\bm v}},T)=\frac{m}{2\pi k_{\rm B}T} \exp \left(-\frac{m(\tilde{v}_x^2+\tilde{v}_y^2)}{2k_{\rm B} T}\right),\label{eq.mb_distri}
\end{eqnarray}
where $T$ can be regarded as the molecular kinetic temperature defined by the averaged kinetic energy per degree of freedom measured in the moving frame as
\begin{eqnarray}
\frac{k_{\rm B}T}{2}\equiv \int \frac{m}{2}\tilde{v}_x^2 f_{\rm MB}(\tilde{{\bm v}},T) d\tilde{\bm v}=\int \frac{m}{2}\tilde{v}_y^2 f_{\rm MB}(\tilde{{\bm v}},T) d\tilde{\bm v}.\label{eq.eff_temp}
\end{eqnarray}
Then, as the Jacobian associated with the variable transformation is unity, we obtain the velocity distribution $f({\bm v})$ of the ideal gas particles at position 
${\bm x}$ from $f({\bm v})\equiv f_{\rm MB}(v_x-\bar{v}_x(x), v_y, T)$ as
\begin{eqnarray}
f({\bm v})=\frac{m}{2\pi k_{\rm B}T} \exp \left(-\frac{m((v_x-\bar{v}_x(x))^2+v_y^2)}{2k_{\rm B} T}\right).\label{eq.micro_distri}
\end{eqnarray}
The spatially non-uniform shape of this distribution is remarkable as the temperature and the density of the gas are assumed to be spatially uniform inside the cylinder.
Equation~(\ref{eq.micro_distri}) is expected to recover the ordinary Maxwell--Boltzmann distribution with $T=T^{\rm R}$ in the quasistatic limit $u \to 0$, where the global equilibrium between the working substance and the heat reservoir holds.

\subsection{First law of thermodynamics as time-evolution equation of temperature}\label{sec.1st_law}
We introduce the first law of thermodynamics (the law of energy conservation) 
as a time-evolution equation of the temperature of the gas
by calculating the total energy of the gas, heat flow, and power based on 
Eq.~(\ref{eq.micro_distri}) as follows:
The energy density $e({\bm x})$ of the gas at position $\bm x$ is given by
\begin{eqnarray}
e({\bm x})&&\equiv \frac{N}{V}\int d{\bm v}\frac{m(v_x^2+v_y^2)}{2}f({\bm v})\nonumber\\
&&=\frac{N}{V}k_{\rm B}T+\frac{N}{V}\frac{m}{2}\bar{v}_x(x)^2,
\end{eqnarray}
where $V\equiv Ll$ is the volume of the cylinder.
By performing a spatial integral, we obtain the total energy $E$ of the gas as
\begin{eqnarray}
E=\int_0^l dx \int_0^L dy e({\bm x})=Nk_{\rm B}T+\frac{Nm}{6}u^2.\label{eq.energy}
\end{eqnarray}
The first term is the internal energy $U$ of the 2D ideal gas at temperature $T$, while the second term is the kinetic energy of the fluid.

The heat flow from the thermal wall at the bottom of the cylinder is obtained by using Eq.~(\ref{eq.micro_distri}) at $x=0$~\cite{note}:
\begin{eqnarray}
f({\bm v})|_{x=0}=\frac{m}{2\pi k_{\rm B}T} \exp \left(-\frac{m(v_x^2+v_y^2)}{2k_{\rm B} T}\right).
\end{eqnarray}
By using this distribution, we obtain the following expression of the heat flow according to the procedure developed in~\cite{IO2008}:
We first count the number of particles $n_{\rm in}$ that collide with the thermal wall per unit time as
\begin{eqnarray}
n_{\rm in}&& \equiv \int_{-\infty}^{0}dv_x \int_{-\infty}^{\infty}dv_y\frac{N}{V}L_{\rm th}(-v_x) f({\bm v})|_{x=0}\nonumber\\
&&=\frac{L_{\rm th}N}{2\pi V}\sqrt{\frac{2\pi k_{\rm B}T}{m}}.
\end{eqnarray}
The energy $q_{\rm in}$ flowing from the colliding particles into the thermal wall per unit time is also calculated as
\begin{eqnarray}
q_{\rm in}&&\equiv \int_{-\infty}^0dv_x \int_{-\infty}^{\infty}dv_y \frac{N}{V}L_{\rm th}\frac{m(v_x^2+v_y^2)}{2}(-v_x)f({\bm v})|_{x=0}\nonumber\\
&&=\frac{3L_{\rm th}Nk_{\rm B}T}{4\pi V}\sqrt{\frac{2\pi k_{\rm B}T}{m}}.
\end{eqnarray}
Because the number of the reflected particles $n_{\rm out}$ per unit time should be equal to $n_{\rm in}$, we can calculate the energy flowing into the working substance as
\begin{eqnarray}
q_{\rm out}&&\equiv n_{\rm in}\int_0^{\infty}dv'_x \int_{-\infty}^{\infty}dv'_y\frac{m({v_x'}^2+{v'_y}^2)}{2} f_{\rm th}({\bm v}')\nonumber\\
&&=\frac{3L_{\rm th}Nk_{\rm B}T^{\rm R}}{4\pi V}\sqrt{\frac{2\pi k_{\rm B}T}{m}}
\end{eqnarray}
by using Eq.~(\ref{eq.distri_thermal}).
Then the heat flow $q\equiv q_{\rm out}-q_{\rm in}$ is obtained as
\begin{eqnarray}
q=\sqrt{\frac{2\pi k_{\rm B}T}{m}}\frac{3L_{\rm th}Nk_{\rm B}(T^{\rm R}-T)}{4\pi V}\equiv \kappa (T^{\rm R}-T),\label{eq.fourier}
\end{eqnarray}
where we have defined the following thermal conductance 
\begin{eqnarray}
\kappa \equiv \sqrt{\frac{2\pi k_{\rm B}T}{m}}\frac{3L_{\rm th}Nk_{\rm B}}{4\pi V(t)}.\label{eq.thermal_conduc}
\end{eqnarray}
This depends on time $t$ through the volume change even if the temperature $T$ does not change with time.
Therefore, although Eq.~(\ref{eq.fourier}) has the form of the linear Fourier's law of heat transport, it is different from the setup in the original CA cycle,
where $\kappa$ is assumed 
not to depend on $T$ and $t$~\cite{CA}.

To calculate the power as the work done on the piston per unit time, we use Eq.~(\ref{eq.micro_distri}) at $x=l$:
\begin{eqnarray}
f({\bm v})|_{x=l}=\frac{m}{2\pi k_{\rm B}T} \exp \left(-\frac{m((v_x-u)^2+v_y^2)}{2k_{\rm B} T}\right).
\end{eqnarray}
Then the power $w$ is calculated by using this distribution and the work per collision Eq.~(\ref{eq.micro_work}) as
\begin{eqnarray}
w&&=\int_{-\infty}^{\infty}dv_y\int_{u}^{\infty}d{v}_x 2mu (v_x-u)^2\frac{N}{V}Lf({\bm v})|_{x=l}\nonumber\\
&&=\frac{Nk_{\rm B}T}{V}Lu=p\frac{dV}{dt},\label{eq.work_flux}
\end{eqnarray}
which is expressed by the product of the pressure and the time derivative of the volume, where we used the equation of state for the ideal gas $p=\frac{Nk_{\rm B}T}{V}$.

By using Eqs.~(\ref{eq.energy}), (\ref{eq.fourier}), and (\ref{eq.work_flux}), we finally obtain the first law of thermodynamics 
$\frac{dU}{dt}=Nk_{\rm B}\frac{dT}{dt}=q-w$ as
\begin{eqnarray}
Nk_{\rm B}\frac{dT(t)}{dt}=\kappa(t) (T^{\rm R}-T(t))-\frac{Nk_{\rm B}T(t)}{V(t)}\frac{dV(t)}{dt},\label{eq.1st_law_deriv}
\end{eqnarray}
which serves as the time-evolution equation of the temperature of the gas.
We can also validate Eq.~(\ref{eq.1st_law_deriv}) based on 
the inviscid Navier-Stokes equations 
(see the Appendix~\ref{appendix}).
We note that reducing the dynamics of the gas 
into the time-evolution equation of the spatially uniform temperature 
in this way is an approximation based on the separation of the time scales 
(see the last paragraph in Sec.~\ref{Ideal gas system in cylinder with moving piston}),
where validity of the results obtained under this approximation should be verified by a molecular dynamics simulation.

\section{Local equilibrium Carnot cycle}\label{Local equilibrium Carnot cycle}
\begin{figure}
\begin{center}
\includegraphics[scale=0.75]{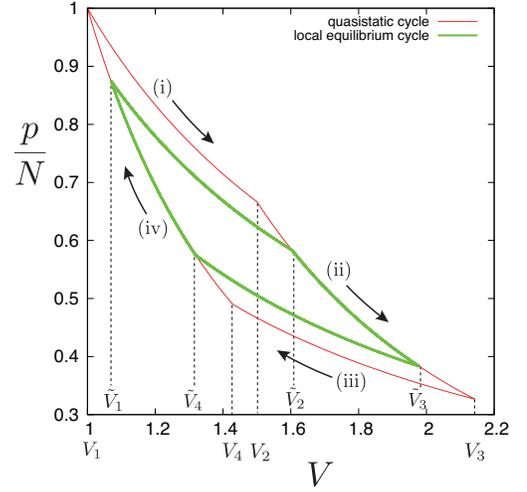}
\end{center}
\caption{(Normalized) pressure--volume ($p$--$V$) diagram of the local equilibrium Carnot cycle of the 2D ideal gas
under the parameters $N=100$, $k_{\rm B}=m=L=1$, $T_h^{\rm R}=1$, $T_c^{\rm R}=0.7$, $V_1=1$, $V_2=1.5$, $L_{{\rm th},h}=L_{{\rm th},c}=0.05$, and $u_h=-u_c=2\times 10^{-3}$.
The thin curve represents the quasistatic (global equilibrium) cycle. The bold curve represents the local equilibrium cycle, where $p=\frac{Nk_{\rm B}T_i^{\rm st}}{V}$ with Eq.~(\ref{eq.steady_temp2}) during the isothermal processes and Eq.~(\ref{eq.p_v_adi}) during the adiabatic processes.
$\tilde{V}_j$ denotes the switching volume of the local equilibrium Carnot cycle as in Eq.~(\ref{eq.switch_finite}) and $V_j$ denotes the corresponding volume of the quasistatic cycle.}
\label{t_v}
\end{figure}

\subsection{Construction of cycle}
We construct the local equilibrium Carnot cycle of the ideal gas based on the preparation in Sec.~\ref{Molecular kinetic model}. 
Hereafter, the suffix $i$ ($i=h,c$) denotes the quantity during the isothermal processes in contact with the heat reservoir with the temperature $T_i^{\rm R}$.

We require that the local equilibrium Carnot cycle should recover the quasistatic Carnot cycle in the quasistatic limit $u_i \to 0$. 
We denote by $V_j$ ($j=1, \cdots,4$) the volume at which we switch each thermodynamic process of the quasistatic cycle.
The quasistatic Carnot cycle consists of the following successive thermodynamic processes (Fig.~\ref{t_v}):
(i) the isothermal expansion process in contact with the heat reservoir with $T_h^{\rm R}$ (${V}_1\to {V}_2$); (ii) the adiabatic expansion process (${V}_2\to {V}_3$); (iii) the isothermal compression process (${V}_3\to {V}_4$) in contact with the heat reservoir with $T_c^{\rm R}$; (iv) the adiabatic compression process (${V}_4\to {V}_1$).
Because the adiabatic equation of the $2$D ideal gas $TV={\rm const}.$ holds for the quasistatic adiabatic process, $V_j$'s depend on each other as
\begin{eqnarray}
V_3=\frac{T_h^{\rm R}}{T_c^{\rm R}}V_2, \ 
V_4=\frac{T_h^{\rm R}}{T_c^{\rm R}}V_1,\label{eq.qs_vol}
\end{eqnarray}
showing that the independent variables are only $V_1$ and $V_2$ when we fix the temperatures $T_h^{\rm R}$ and $T_c^{\rm R}$.

Denoting by $\tilde{V}_j$ the volume at which we switch each thermodynamic process 
depending on the constant piston velocity 
and 
defining the cylinder length $\tilde{l}_j$ at the switching volume as $\tilde l_j \equiv {\tilde V}_j/L$,
we design our local equilibrium cycle consisting of the successive thermodynamic processes as follows (Fig.~\ref{t_v}): (i) the isothermal expansion process with piston velocity $u_h$ in contact with the heat reservoir with $T_h^{\rm R}$ ($\tilde{V}_1\to \tilde{V}_2$)
[the duration of this process is $t_h\equiv ({\tilde{l}_2-\tilde{l}_1})/{u_h}$, and
the temperature of the working substance always takes the steady value $T_h^{\rm st}\equiv T_h(u_h)$ ($\le T_h^{\rm R}$)]; (ii) the adiabatic expansion process
with duration $\gamma t_h$
($\tilde{V}_2\to \tilde{V}_3$); (iii) the isothermal compression process with piston velocity $u_c$ 
in contact with the heat reservoir with $T_c^{\rm R}$ ($\tilde{V}_3\to \tilde{V}_4$)
[the duration of this process is $t_c\equiv ({\tilde{l}_4-\tilde{l}_3})/{u_c}$, and
the temperature of the working substance always takes the steady value $T_c^{\rm st}\equiv T_c(u_c)$ ($\ge T_c^{\rm R}$)]; 
(iv) the adiabatic compression process with duration $\gamma t_c$
($\tilde{V}_4\to \tilde{V}_1$).
In this design, the total duration completing the adiabatic processes is proportional to $t_h+t_c$, as assumed in~\cite{CA}.
While there may be many ways of switching each process for $\tilde{V}_j$ to recover $V_j$ in the quasistatic limit $u_i\to 0$, 
we adopt the following switching volumes depending on $u_i$ through $T_i^{\rm st}$ as
\begin{eqnarray}
\tilde{V}_1=\frac{T_h^{\rm R}}{T_h^{\rm st}}V_1, \tilde{V}_2=\frac{T_h^{\rm R}}{T_h^{\rm st}}V_2, \tilde{V}_3=\frac{T_c^{\rm R}}{T_c^{\rm st}}V_3, \tilde{V}_4=\frac{T_c^{\rm R}}{T_c^{\rm st}}V_4.\label{eq.switch_finite} 
\end{eqnarray}
Because, as shown below, the adiabatic equation $TV={\rm const}.$ holds irrespective of $u_i$,
the adiabatic processes of the local equilibrium cycle as switched by Eq.~(\ref{eq.switch_finite}) always overlap with the quasistatic adiabatic ones (see Fig.~\ref{t_v})~\cite{IO2015}, and they end with the steady temperatures of the succeeding isothermal processes.

To obtain the steady temperature $T_i^{\rm st}$, 
we consider the time-evolution equation of the gas in Eq.~(\ref{eq.1st_law_deriv}) 
during the isothermal processes: 
\begin{eqnarray}
Nk_{\rm B}\frac{dT_i(t)}{dt}=\kappa_i(t) (T_i^{\rm R}-T_i(t))-\frac{Nk_{\rm B}T_i(t)}{V(t)}\frac{dV(t)}{dt}.\label{eq.1st_law}
\end{eqnarray}
We can obtain the steady solution $T^{\rm st}_i$ of Eq.~(\ref{eq.1st_law}) that satisfies $\frac{dT_i(t)}{dt}=0$, solving a quadratic equation in $T^{\rm st}_i$,
\begin{eqnarray}
T_i^{\rm R}-T_i^{\rm st}=\frac{u_i}{A_i}\sqrt{T_i^{\rm st}},\label{eq.steady_temp} 
\end{eqnarray}
where we used Eq.~(\ref{eq.thermal_conduc}) and 
\begin{eqnarray}
A_i\equiv \sqrt{\frac{2\pi k_{\rm B}}{m}}\frac{3L_{{\rm th},i}}{4\pi L},\label{eq.A}
\end{eqnarray} 
where we consider heat-reservoir dependence of $L_{\rm th}$, which also leads to
heat-reservoir dependence of the thermal conductance $\kappa$ in Eq.~(\ref{eq.thermal_conduc}) as~\cite{CA}.
The solution of Eq.~(\ref{eq.steady_temp}) is given by
\begin{eqnarray}
T^{\rm st}_i=T_i^{\rm R}-\frac{u_i}{2A_i}\sqrt{4T_i^{\rm R}+\frac{u_i^2}{A_i^2}}+\frac{u_i^2}{2A_i^2},\label{eq.steady_temp2}
\end{eqnarray}
where we have chosen the minus sign as a physically relevant solution.
In the quasistatic limit $u_i \to 0$, we can see that $T^{\rm st}_i$ agrees with $T_i^{\rm R}$ of the heat reservoir, as expected.
This exact relation between the steady temperature and the piston velocity is a merit obtained by 
our microscopic formulation using the specific working substance, which cannot be obtained by general but phenomenological approaches~\cite{CA,IO2015}.

Since the first term on the right-hand side of Eq.~(\ref{eq.1st_law}) vanishes in the adiabatic
processes, we obtain the adiabatic relation between $T$ and $V$ as $TV={\rm const}.$ 
holding irrespective of $u_i$, by directly solving Eq.~(\ref{eq.1st_law}). From this we can
validate Eq.~(\ref{eq.switch_finite}).
The relation
\begin{eqnarray}
pV^2=\rm const.\label{eq.p_v_adi}
\end{eqnarray}
also follows from the equation of state $p=\frac{Nk_{\rm B}T}{V}$ as is used to depict the adiabatic curves 
in Fig.~\ref{t_v}.

The entropy of the ideal gas with temperature $T$ is calculated by using $f({\bm v})$ in Eq.~(\ref{eq.micro_distri}) from the Gibbs entropy formula as
\begin{eqnarray}
S(T,V)&&=-Nk_{\rm B}\int \frac{f({\bm v})}{V}\ln \frac{f({\bm v})}{V}d{\bm v}d{\bm x}\nonumber\\
&&=Nk_{\rm B}\ln T+Nk_{\rm B}\ln V+S_0,\label{eq.entropy}
\end{eqnarray}
where $S_0$ is a constant independent of $T$ and $V$.
It is then easy to confirm 
\begin{eqnarray}
q_i(t)=\kappa_i(t)(T_i^{\rm R}-T_i^{\rm st})=T_i^{\rm st}\frac{dS}{dt}\label{eq.heat_entropy_exp}
\end{eqnarray}
from Eqs.~(\ref{eq.thermal_conduc}), (\ref{eq.steady_temp}) and (\ref{eq.entropy}). 
From this definition of the entropy, 
it turns out that the switching volumes in Eq.~(\ref{eq.switch_finite})
maintain the entropy change during the isothermal process at any piston velocity $u_i$ as 
$\Delta S_h \equiv Nk_{\rm B}\ln \frac{\tilde{V}_2}{\tilde{V}_1}=Nk_{\rm B}\ln \frac{{V}_2}{{V}_1}\equiv \Delta S$ and $\Delta S_c \equiv Nk_{\rm B}\ln \frac{\tilde{V}_4}{\tilde{V}_3}=Nk_{\rm B}\ln \frac{{V}_4}{{V}_3}=-\Delta S$, where we used Eq.~(\ref{eq.qs_vol})~\cite{IO2015}.

\subsection{Efficiency and power}
The net heat from the heat reservoir during each isothermal process is calculated as
\begin{eqnarray}
Q_i=\int_0^{t_i}\kappa_i(t)(T_i^{\rm R}-T^{\rm st}_i)dt=T_i^{\rm st}\Delta S_i,\label{eq.heat_avr}
\end{eqnarray}
where we used Eq.~(\ref{eq.heat_entropy_exp}).
This is the local-equilibrium counterpart of the quasistatic heat $T_i^{\rm R}\Delta S_i$
with $T_i^{\rm R}$ being replaced with $T_i^{\rm st}$ of the working substance~\cite{IO2015}. 
From Eq.~(\ref{eq.heat_avr}) and $\Delta S_h=-\Delta S_c=\Delta S$, the efficiency of the present local equilibrium Carnot cycle is given by 
\begin{eqnarray}
&&\eta=1+\frac{Q_c}{Q_h}=1-\frac{T_c^{\rm st}}{T_h^{\rm st}},\label{eq.effi}
\end{eqnarray}
which corresponds to the endoreversible expression of Eq.~(\ref{eq.endo_effi}) in the present model, revealing that $T_i$ in Eq.~(\ref{eq.endo_effi}) is 
the steady value of the molecular kinetic temperature of the working substance as defined in Eq.~(\ref{eq.eff_temp}). 

By using Eqs.~(\ref{eq.qs_vol}), (\ref{eq.switch_finite}), and (\ref{eq.steady_temp}), we can express the power of our cycle by using $T_i^{\rm st}$ without $u_i$ as~\cite{CA}
\begin{align}
P&\equiv \frac{W}{(1+\gamma)(t_h+t_c)}=\frac{(T_h^{\rm st}-T_c^{\rm st})\Delta S}{(1+\gamma)\left(\frac{{\tilde{l}_2-\tilde{l}_1}}{{u_h}}+\frac{{\tilde{l}_4-\tilde{l}_3}}{{u_c}}\right)}\nonumber\\
&=\frac{A_hA_c\Delta S \sqrt{(T_c^{\rm R}-y)(T_h^{\rm R}-x)}(\Delta T^{\rm R}-x+y)xy}{(1+\gamma)(l_2-l_1)T_h^{\rm R} \left(A_c y\sqrt{T_c^{\rm R}-y}-A_h x \sqrt{T_h^{\rm R}-x}\right)},
\label{eq.pow_full}
\end{align}
where we have defined $x \equiv T_h^{\rm R}-T_h^{\rm st}$, $y \equiv T_c^{\rm R}-T_c^{\rm st}$, and $\Delta T^{\rm R}\equiv T_h^{\rm R}-T_c^{\rm R}$.

\subsection{Efficiency at maximum power}
In principle, by maximizing the power Eq.~(\ref{eq.pow_full}) as $\frac{\partial P}{\partial x}=\frac{\partial P}{\partial y}=0$ as done in the original CA paper~\cite{CA}, we can obtain the efficiency at maximum power of our cycle. 
This is, however, difficult to perform analytically in general. 
Therefore, we focus here on the case of a small temperature difference $\Delta T^{\rm R}$ for this analytic treatment as a guideline.
In this case,
we obtain the power instead of Eq.~(\ref{eq.pow_full}) as
\begin{eqnarray}
P=\frac{A_hA_c\Delta S}{(1+\gamma)(l_2-l_1)\sqrt{\bar{T}^{\rm R}}}\frac{(\Delta T^{\rm R}-x+y)xy}{A_cy-A_hx},\label{eq.pow_approx}
\label{eq.pow}
\end{eqnarray}
to the lowest order of $\Delta T^{\rm R}$, $x$ and $y$, where $\bar{T}^{\rm R}\equiv (T_h^{\rm R}+T_c^{\rm R})/2$.
By maximizing the power Eq.~(\ref{eq.pow}) as $\frac{\partial P}{\partial x}=\frac{\partial P}{\partial y}=0$,
we easily obtain the $x$ and $y$ values at maximum power as
\begin{eqnarray}
x^*=\frac{\sqrt{A_c}\Delta T^{\rm R}}{2(\sqrt{A_h}+\sqrt{A_c})}, \ y^*=-\frac{\sqrt{A_h}\Delta T^{\rm R}}{2(\sqrt{A_h}+\sqrt{A_c})}.\label{eq.temp_at_maxpow}
\end{eqnarray}
Then the maximum power and the efficiency at maximum power turn out to be
\begin{eqnarray}
&&P^*=\frac{A_hA_c\Delta S}{4(1+\gamma)(l_2-l_1)\sqrt{\bar{T}^{\rm R}}}\frac{{\Delta T^{\rm R}}^2}{(\sqrt{A_h}+\sqrt{A_c})^2},\label{eq.max_pow}\\
&&\eta^*=1-\frac{T_c^R-y^*}{T_h^R-x^*}=\frac{\eta_{\rm C}}{2-\frac{\eta_{\rm C}}{1+\sqrt{\frac{A_h}{A_c}}}},\label{eq.ss}
\end{eqnarray}
respectively.
This expression of $\eta^*$ is essentially the same as the Schmiedl--Seifert efficiency in a stochastic heat engine model~\cite{SS2008}.
By expanding Eq.~(\ref{eq.ss}) with respect to $\eta_{\rm C}$, we obtain 
\begin{eqnarray}
\eta^*=\frac{\eta_{\rm C}}{2}+\frac{\eta_{\rm C}^2}{4\left(1+\sqrt{\frac{A_h}{A_c}}\right)}+O(\eta_{\rm C}^3).
\end{eqnarray}
The linear order agrees with that of the CA efficiency Eq.~(\ref{eq.ca}),
which has been shown to be the upper bound of $\eta^*$ in the linear response regime~\cite{VB}.
This bound is attained by heat engines with the tight-coupling property between the heat and the motion fluxes 
without heat-leakage~\cite{VB}, 
which is satisfied in our present model.
The quadratic order also recovers that of the CA efficiency Eq.~(\ref{eq.ca})
under the symmetric condition of $A_h=A_c$, i.e., $L_{{\rm th},h}=L_{{\rm th},c}$ from Eq.~(\ref{eq.A})~\cite{ELV}. 
The efficiency of the same form as Eq.~(\ref{eq.ss}) has also previously been obtained such as in 
the low-dissipation Carnot cycle~\cite{EKLV}, 
the minimally nonlinear irreversible heat engine~\cite{IO2012}, and the heat engine based on the weighted thermal flux~\cite{ST},
which describe heat engines to the lowest degree of nonequilibrium from the quasistatic limit.

The reason why we have obtained Eq.~(\ref{eq.ss}) rather than the CA efficiency Eq.~(\ref{eq.ca}) can be considered as follows:
A crucial difference between our model and the CA model is that the steady temperature during the isothermal process Eq.~(\ref{eq.steady_temp2}) as a function of the piston velocity 
is available owing to the time-evolution equation Eq.~(\ref{eq.1st_law}) in our case.
Because the approximation Eq.~(\ref{eq.pow_approx}) is equivalent to considering only the lowest correction to the quasistatic limit in Eq.~(\ref{eq.steady_temp2}) as $T_i^{\rm st}\simeq T_i^{\rm R}-\frac{u_i}{A_i}\sqrt{T_i^{\rm R}}$ 
together with the quasistatic-case switching volumes Eq.~(\ref{eq.qs_vol}) for a small temperature difference $\Delta T^{\rm R}$,
it is natural that it yields the efficiency like Eq.~(\ref{eq.ss}) as similar to the other models~\cite{EKLV,IO2012,ST} rather than the CA efficiency.
As the temperature difference increases, we expect that 
the higher-order terms of the piston velocity in Eq.~(\ref{eq.steady_temp2}) together with the piston-velocity dependent switching volumes Eq.~(\ref{eq.switch_finite}) 
that are not adopted in the other models may give rise to a discrepancy between our model and the other models. 

\begin{figure}
\begin{center}
\includegraphics[scale=0.75]{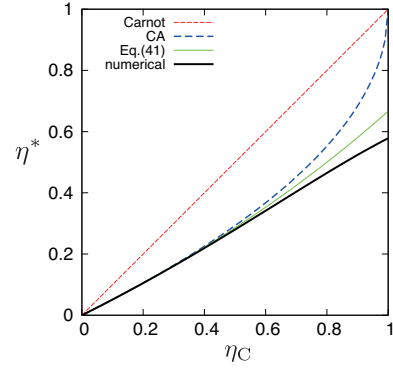}
\caption{Efficiency at maximum power $\eta^*$ under the symmetric condition of $A_h=A_c$ as a function of $\eta_{\rm C}=1-T_c^{\rm R}$, with $T_h^{\rm R}=1$. 
The numerical curve indicates $\eta^*$ obtained by maximizing Eq.~(\ref{eq.pow_full}) with respect to $x$ and $y$ 
numerically.}
\label{effi_at_pmax}
\end{center}
\end{figure}

In Fig.~\ref{effi_at_pmax}, we show $\eta^*$
obtained by maximizing Eq.~(\ref{eq.pow_full}) with respect to $x$ and $y$ 
numerically and the analytical result Eq.~(\ref{eq.ss}) in the case of $A_h=A_c$. 
The CA efficiency in Eq.~(\ref{eq.ca}) is also shown for comparison.
We can confirm that the numerical value agrees with Eq.~(\ref{eq.ss}) and the CA efficiency for the small temperature difference, 
while it begins to deviate from these efficiencies as the temperature difference increases.

\subsection{Verification by molecular dynamics simulation}

To verify the validity of our theory, we performed an event-driven molecular dynamics (MD) simulation~\cite{AW} 
of our local equilibrium Carnot cycle by regarding the 2D ideal gas particles as low-density hard discs~\cite{IO2008} with diameter $d$.

\begin{figure}
\begin{center}
\includegraphics[scale=0.65]{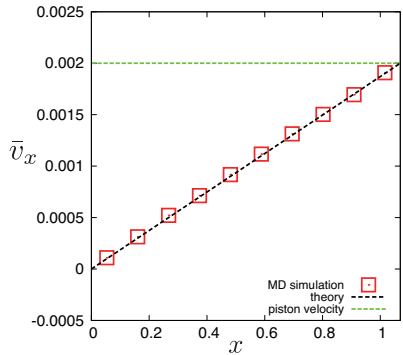}
\end{center}
\caption{Local center-of-mass $x$ velocity of the gas particles
obtained from an MD simulation. 
The same parameters as in Fig.~\ref{t_v} are used with $\gamma=0.5$, $d=0.01$, $l_m=\tilde{l}_1\simeq 1.069$, $\Delta l=0.1$, and $N_{\rm cell}=10$.
We used $262400$ cycles for the average (see the main text).
The theoretical line is given by Eq.~(\ref{eq.local_velocity}) with $l=l_m$.}
\label{center_velocity}
\end{figure}

In Fig.~\ref{center_velocity}, we show the local center-of-mass $x$-velocity $\bar{v}_x(x_k)$
obtained from the simulation as follows:
When the cylinder length is $l_m < l < l_m+\Delta l$ during the isothermal expansion processes, where $l_m$ is the starting point of measurement and $\Delta l$ is a small displacement, 
we divide the cylinder $l \times L$
into small cells $X_k \times L$ in the $x$-direction with $X_k=[X_k^{\rm min}, X_k^{\rm max}]\equiv \left[\frac{l}{N_{\rm cell}}(k-1), \frac{l}{N_{\rm cell}}k\right]$ ($k=1, \cdots, N_{\rm cell}$). 
At every particle--particle collision event that occurs during $l_m < l < l_m+\Delta l$ along repeated cycles,
we measure the $x$ velocity of the particles belonging to each cell.
We define the local center-of-mass $x$ velocity at the $k$th cell $\bar{v}_x(x_k)$ as the average of all the $x$ velocities measured in the $k$th cell,
where $x_k \equiv \frac{X_k^{\rm min}+X_k^{\rm max}}{2}$.
We can see that $\bar{v}_x(x_k)$ agrees with the theoretical line Eq.~(\ref{eq.local_velocity}) well.

\begin{figure}
\includegraphics[scale=0.63]{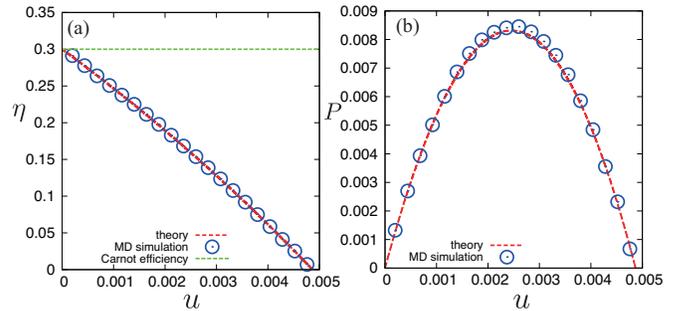}
\caption{(a) Efficiency and (b) power as functions of $u_h=-u_c\equiv u$. 
The same parameters as in Fig.~\ref{t_v}, except the piston velocity are used with $\gamma=0.5$ and $d=0.01$.
We used 3200--76160 cycles for the average. The theoretical Carnot efficiency is $\eta_{\rm C}=0.3$.}
\label{efficiency_power}
\end{figure}

In Fig.~\ref{efficiency_power}, we also compare the efficiency and power obtained by summing the heat and work per collision Eqs.~(\ref{eq.micro_heat}) and (\ref{eq.micro_work}) 
by an MD simulation with the theoretical values Eqs.~(\ref{eq.effi}) and (\ref{eq.pow_full}) using $T_i^{\rm st}$ in Eq.~(\ref{eq.steady_temp2}) for the case of $u_h=-u_c$, which show a good agreement over the whole working regime.

\section{Discussion and summary}\label{Discussion and summary}
We previously studied a finite-time Carnot cycle of 2D ideal gas~\cite{IO2008} based on molecular kinetics  
in a setup similar to that in the present work. 
Although in that work we used a usual Maxwell--Boltzmann distribution with a well-defined temperature $T$
of the gas as the velocity distribution of the particles,
it was just an assumption without considering the spatial variation of the distribution.
Because of the lack of this spatial variation, the fundamental thermodynamic relation as Eq.~(\ref{eq.ftr}) did not hold for the model in~\cite{IO2008}.
Moreover, we constructed the finite-time Carnot cycle by switching each thermodynamic process at the same volumes as in the quasistatic cycle. 
This led to an extra heat transfer for relaxation of the working substance to a steady temperature during the isothermal processes, 
which do not exist in the original CA cycle~\cite{CA}.
In the present local equilibrium Carnot cycle, we have overcome these difficulties in~\cite{IO2008} by deriving the velocity distribution with reasonable spatial variation Eq.~(\ref{eq.micro_distri}),
and by appropriately switching each thermodynamic process depending on the piston velocity so that such an extra heat transfer does not occur.

In the present paper, we identified the velocity distribution Eq.~(\ref{eq.micro_distri}) of 2D ideal gas as the working substance that is compatible with the local equilibrium assumption and the fundamental thermodynamic relation
satisfying the endoreversibility. 
We found that this distribution 
is the Maxwell--Boltzmann distribution with the spatially uniform temperature and the
spatially varying local center-of-mass velocity Eq.~(\ref{eq.local_velocity}).
Based on this distribution, we obtained the time-evolution equation of the temperature of the gas. 
We then constructed the local equilibrium Carnot cycle by using the steady solution of the equation.
We confirmed that the efficiency of the present local equilibrium Carnot cycle is given by the endoreversible Carnot efficiency using the steady values of the molecular kinetic temperatures of the working substance during the isothermal processes.
We also studied the efficiency at maximum power of our cycle, and showed that it is given by the Schmiedl--Seifert efficiency~\cite{SS2008} under a small temperature difference. 
We have numerically confirmed the local center-of-mass velocity Eq.~(\ref{eq.local_velocity}) by performing an MD simulation.
We expect that our theory gives a nonequilibrium statistical mechanics basis for the endoreversible heat engines and finite-time thermodynamics.

\acknowledgments
Y. I. acknowledges the financial support from JSPS KAKENHI Grant No. 16K17765.

\appendix
\section{Consistency with inviscid Navier-Stokes equations}\label{appendix}
We validate the local center-of-mass $x$ velocity Eq.~(\ref{eq.local_velocity}) 
derived in Sec.~\ref{sec.local_vel} and the first law of thermodynamics Eq.~(\ref{eq.1st_law_deriv}) derived in Sec.~\ref{sec.1st_law}
based on the following fluid-mechanical argument:
Dynamics of a 2D compressible inviscid fluid is determined by the following mass-, momentum-, and energy-conservation equations~\cite{EM}
\begin{eqnarray}
&&\frac{\partial \rho}{\partial t}+\nabla \cdot (\rho \bar{\bm v})=0,\label{eq.mass}\\
&&\frac{\partial (\rho \bar{\bm v})}{\partial t}+\nabla \cdot (\rho \bar{\bm v}\bar{\bm v})+\nabla p={\bm 0,}\label{eq.momentum}\\
&&\frac{\partial e}{\partial t}+\nabla \cdot \left((e+p)\bar{\bm v}+{\bm J}\right)=0,\label{eq.energy_density}
\end{eqnarray}
respectively.
Here, $\bar{\bm v}({\bm x},t)$ is the fluid velocity corresponding to our local center-of-mass velocity, $\rho({\bm x},t)$ is the mass density, $p({\bm x},t)$ is the pressure, $e({\bm x},t)$
is the energy density, and ${\bm J}({\bm x},t)$ is the heat flux.
To be more specific, the fluid is a 2D ideal gas with $p({\bm x},t)=\frac{\rho({\bm x},t)}{m}k_{\rm B}T({\bm x},t)$ and $e({\bm x},t)=p({\bm x},t)+\frac{1}{2}\rho({\bm x},t)\bar{{\bm v}}^2({\bm x},t)$,
where $p({\bm x},t)$ serves as the internal energy density of the 2D ideal gas.
We assume that the fluid is uniform in the $y$-direction and the $y$-component 
of the fluid velocity vanishes as
$\bar{\bm v}({\bm x},t)=(\bar{v}_x(x,t), 0)$.
Equation~(\ref{eq.momentum}) can then be reduced to a $1$D inviscid Navier-Stokes equation:
\begin{eqnarray}
&&\frac{\partial \bar{v}_x}{\partial t}+\bar{v}_x\frac{\partial \bar{v}_x}{\partial x}+\frac{1}{\rho}\frac{\partial p}{\partial x}=0,\label{eq.momentum_2}
\end{eqnarray}
where we used Eq.~(\ref{eq.mass}).
By assuming a spatially uniform mass density and temperature as the endoreversibility condition as
\begin{eqnarray}
&&\rho({\bm x},t)=\rho(t)=\frac{mN}{Ll(t)},\label{eq.endo_1}\\
&&T({\bm x},t)=T(t),\label{eq.endo_2}
\end{eqnarray} 
we can directly solve Eq.~(\ref{eq.momentum_2}) as follows: The separation of variables $\bar{v}_x(x,t)=F(x)G(t)$ yields 
\begin{eqnarray}
\frac{dF}{dx}=u,\label{eq.f}\\
-\frac{1}{G^2}\frac{dG}{dt}=u,\label{eq.g}
\end{eqnarray}
where $u$ is an arbitrary constant independent of $x$ and $t$. By solving Eqs.~(\ref{eq.f}) and (\ref{eq.g}), 
we obtain $F(x)=ux+C_1$ and $G(t)=\frac{1}{ut+C_2}$, where $C_1$ and $C_2$ are integral constants.
By imposing $F(0)=0$ and $G(0)=\frac{1}{l_0}$, we obtain
\begin{eqnarray}
\bar{v}_x(x,t)=\frac{ux}{ut+l_0},\label{eq.fluid_vel}
\end{eqnarray}
which agrees with the local center-of-mass velocity Eq.~(\ref{eq.local_velocity}) by regarding $u$ and $l_0$ as the 
constant piston velocity and the initial position of the piston, respectively.
We can confirm that $x$ component Eq.~(\ref{eq.fluid_vel}) and vanishing $y$ component of the fluid velocity also satisfy Eq.~(\ref{eq.mass}).

We next consider the energy conservation equation Eq.~(\ref{eq.energy_density}).
From the endoreversibility condition Eqs.~(\ref{eq.endo_1}) and (\ref{eq.endo_2}) and Eq.~(\ref{eq.fluid_vel}), Eq.~(\ref{eq.energy_density}) becomes
\begin{eqnarray}
\frac{Nk_{\rm B}}{V}\frac{dT}{dt}=-\nabla \cdot {\bm J}-\frac{Nk_{\rm B}T}{Vl} \frac{dl}{dt}.\label{eq.energy_2}
\end{eqnarray}
By performing a spatial integral on both sides of this equation, 
we obtain
\begin{eqnarray}
Nk_{\rm B}\frac{dT}{dt}=q-p\frac{dV}{dt},\label{eq.1st_law_fluid}
\end{eqnarray}
where we defined $q\equiv \int_0^L J_x(0,y,t)dy$ and 
used $J_y(x,0,t)=J_y(x,L,t)=J_x(l,y,t)=0$ except at
the thermal wall of the cylinder.
Equation~(\ref{eq.1st_law_fluid}) corresponds to the first law of thermodynamics Eq.~(\ref{eq.1st_law_deriv}),
where the detailed form of $q$ has been determined by the molecular kinetics.


\begin{references}
\bibitem{note} Even when the thermal wall is put on a side of the cylinder instead of the bottom of the cylinder, as in the present setup,
the local center-of-mass $x$ velocity [Eq.~(\ref{eq.local_velocity})] and the velocity distribution 
[Eq.~(\ref{eq.micro_distri})] are not affected by this change as far as the piston moves in the $x$ direction, as is clear from their derivation. 
However, the heat flow in this case is not the exact Fourier type as Eq.~(\ref{eq.fourier}), but has an extra correction term of $O(u^2)$.
\end{references}
\end{document}